# Single universal curve for α decay derived from semi-microscopic calculations

M. Ismail [1], W. M. Seif [1,*], A. Y. Ellithi[1], and A. Abdurrahman[2]

[1] Cairo University, Faculty of Science, Department of Physics, Giza, Egypt
[2] Misr University for Science and Technology (MUST), Faculty of Engineering, Department of Physics, Giza, Egypt

**ABSTRACT**

A universal curve is one of the simple ways to get preliminary information about the α-decay half-life times of heavy nuclei. We try to find parameterization for the universal curve of α decay based on semi-microscopic calculations, starting from the realistic M3Y-Reid nucleon-nucleon interaction. Within the deformed density-dependent cluster model, the penetration probability and the assault frequency are calculated using the Wentzel-Kramers-Brillouin (WKB) approximation. The deformations of daughter nuclei and the ground-state spin and parity of the involved nuclei are considered. For all studied decays, we found that it is accurate enough to express the assault frequency either as a function of the mass number of the parent nuclei or as a constant average value. The average preformation probability of the α cluster inside four groups of 166 even (Z)-even (N), 117 odd-even, 141 even-odd, and 72 odd-odd α-emitters are obtained, individually. The effects of participating unpaired nucleons in the involved nuclei, as well as the influence of any differences in their ground-state spin and/or parity, appear in the obtained average values of the preformation probability. We suggested a single universal curve for α decay with only one parameter. This parameter varies according to the classified four groups. It includes the preformation probability and the average assault frequency, in addition to the pairing contribution.

## I. INTRODUCTION

α-decay is one of the prominent decay modes of heavy and synthesized superheavy nuclei [1,2]. The half-lives of the α-decay modes would provide useful information on the detailed structure of the participating nuclei [3,4,5,6,7,8]. Various microscopic and semi-microscopic approaches [9], and phenomenological methods [10,11], were performed to investigate the α-decay process, as a fundamental quantum-tunneling phenomena. For instance, the density dependent cluster model (DDCM) [12], the preformed cluster model [13], different R-matrix approaches [14,15], and the multistep shell model [16] have been used in different studies. The generalized liquid drop model [17] and the universal curve method [18] are used also to study the α decays. In some cases, the different approaches produce decay life times that differ by more than one order of magnitude [5,19,20]. This appears clearly in terms of the extracted value of the preformation probability by different studies. Values in the range from $10^{-5}$ up to unity are obtained, in few cases, for the preformation probability inside the same nucleus [19, 21, 22], depending on the used model and the considered approximations.

The single universal curve based on the fission approach is a simple method to predict the decay half-life times, for both the α and cluster decay modes, without performing heavy numerical calculations. It is obtained by plotting the sum of the decimal logarithms of the half-life time and the cluster preformation probability versus the decimal logarithm of the barrier penetration probability [18]. A phenomenological penetration probability and constant assault frequency are usually used in the universal curve calculation. Our aim in the present work is to compare the results obtained by the universal curve based on a simple penetration probability with those obtained from more realistic calculations in terms of realistic nucleon-nucleon (NN) interaction. However, we illustrate the validity of representing the realistic calculations performed for the quantities participating in the different stages of the α emission process such as the penetrability and the assault frequency, in the frame work of the DDCM [6,12], by universal curve(s). We consider also the contributions of the different nuclear structure properties, such as the deformations of the involved nuclei and the angular momentum carried by the emitted α particle in the unfavored decays [23].

We organized the layout of the present paper as follows. The next section presents the approximate formulae used in the universal curve calculations. It also outlines our method of calculation. The results and discussion are presented in Sec. III. Finally, the summary and conclusion are given in Sec. IV.

## II. THEORETICAL OUTLINES

A universal curve for α radioactivity can be derived by plotting the sum of the decimal logarithm of the half-life time and that of the corresponding preformation probability versus the decimal logarithm of the penetrability probability through the barrier [18, 24]. It provides a simple mean to estimate rapidly the half-life time ($T_{1/2}$) of a given α decay. As a tunneling process, the partial half-life time of the decay is related to the decay width ($\Gamma$) by,

$$T_{1/2} = \frac{\ln 2}{\Gamma} = \frac{\ln 2}{\nu \, S \, P}. \tag{1}$$

In the simple universal curve calculations, the penetrability probability ($P$) of the preformed α particle through the barrier is usually calculated by the analytical formula,

$$-\log_{10} P_U = 0.22873 (\mu_A Z_e Z_d R_b)^{1/2} \left[ \arccos \sqrt{x} - \sqrt{x}(1-x) \right], \tag{2}$$

where $\mu_A = A_e A_d / (A_e + A_d)$ is the reduced mass number. Here, $R_b$ and $x$ are given in terms of the released energy (Q) at the decay process and the charge (mass) numbers of both the light emitted particle, $Z_e(A_e)$, and daughter nuclei, $Z_d(A_d)$. They read $R_b = 1.43998 \, Z_d Z_e / Q$ and $x = R_t/R_b$, where $R_t = 1.2249 \, (A_d^{1/3} + A_e^{1/3})$. The tunneling assault frequency of the α particle at the barrier ($\nu$) is often considered simply as a fixed value, $\nu = 10^{22.01} S^{-1}$ [25]. Generally, the preformation probability (S) for any emitted light particle of mass number $A_e$ is taken as,

$$\log_{10} S = -0.598 \, (A_e - 1). \tag{3}$$

So, the preformation probability for the α particle inside the parent nucleus becomes $S_\alpha = 0.01607$. Based on Eq.(1), the relation between the half-life time and the penetration probability can be rewritten as [18],

$$log_{10} T_{1/2}(s) = -log_{10} P - log_{10} S + [log_{10}(\ln 2) - log_{10} \nu]. \qquad (4)$$

With approximated constant assault frequency, the third terms on the right hand side is taken, for even-even nuclei, as [24, 25],

$$c_{ee} = log_{10}(\ln 2) - log_{10} \nu = -22.16917. \qquad (5)$$

A hindrance factor ($h_U$), with larger positive values for odd-odd and odd-A nuclei compared with its value for even-even nuclei, is usually added to Eq. (4) to fit the experimental data [24,25]. With Eqs. (3) and (5), the universal curve given by Eq.(4) becomes a straight line. It describes $log_{10} T_{1/2}(s)$ as a function of $log_{10} P$. A single universal curve is obtained for α and cluster decays modes [18]. One can easily calculate $log_{10} P$ from Eq.(2) and simply find $T_{1/2}$ from Eq.(4).

For more realistic calculations, we can use the Wentzel-Kramers-Brillouin (WKB) approximation to find both the two-dimensional penetration probability and the quantum tunneling assault frequency of the tunneled α particle, respectively, as

$$P_{WKB}(\theta) = e^{-2 \int_{R_2(\theta)}^{R_3} k(r,\theta) dr} \qquad (6)$$

$$\nu(\theta) = T^{-1}(\theta) = \left[ \int_{R_1(\theta)}^{R_2(\theta)} \frac{2\mu}{\hbar k(r,\theta)} dr \right]^{-1}, \qquad (7)$$

with the wave number,

$$k(r,\theta) = \sqrt{\frac{2\mu}{\hbar^2} |V_T(r,\theta) - Q|}. \qquad (8)$$

Here, $\mu = \frac{m_e m_d}{m_e + m_d}$ is the reduced mass of the emitted particle and the daughter. $R_i(i = 1,2,3)$ represent three classical turning points along the path of the total interaction potential, $V_T(r,\theta)|_{r=R_i(\theta)} = Q$. One obtains different values for the first two turning points, $R_{i=1,2}(\theta)$, at each emission angle $\theta$. In principle, the deformation of the daughter nucleus plays an important role in the decay process. Compared to spherical daughters, the deformations of daughter nuclei lower the potential barrier at their poles and raise its value at the sides. As a result, this produces a decrease in the value of the calculated half-life time and consequently a decrease in the value of the extracted α-preformation probability. The total interaction potential between the emitted α nucleus and the daughter one is given as the sum of the nuclear ($V_N(r,\theta)$), Coulomb ($V_C(r,\theta)$) and the centrifugal $V_\ell(r)$ terms as,

$$V_T(r,\theta) = \lambda V_N(r,\theta) + V_C(r,\theta) + \frac{(\ell + 1/2)^2 \hbar^2}{2\mu r^2}. \qquad (9)$$

Here, the nuclear contribution of the α-daughter potential is normalized by a scaling factor λ which adjusted to satisfy Bohr-Sommerfeld, or Wildermuth, quantization condition [9]. This quantization condition is necessary in the WKB approximation. The centrifugal part in Eq. (9) is considered in its Langer modified form [12,26,27] to ensure the behavior of the interaction potential near the zero point. We finally evaluate the angel-averaging of the decay width, or its individual components, over the different orientations [21,28]. Both the nuclear and Coulomb [29] parts of the interaction potential, $V_{N(C)}(r,\theta)$, will be obtained from the double folding model,

$$V_{N(C)}(\vec{r}) = \iint \rho_\alpha(\vec{r}_1) \, v_{N(C)}(r_{12}) \rho_D(\vec{r}_2) d\vec{r}_1 d\vec{r}_2. \tag{10}$$

Here, $r_{12}$ is the separation distance between any two interacting nucleons belonging to the two interacting nuclei. $\rho_{\alpha(D)}$ is the density distribution of the $\alpha$ (daughter) nucleus. For the nuclear (Coulomb) part of the interaction potential, the matter (charge) density distributions will be used. $v_C$ is the standard proton-proton Coulomb interaction. The M3Y-Reid type of the NN force, with zero range exchange contribution, has the form [30,31],

$$v_N(r_{12}, E) = v_{00}(r_{12}) + \hat{j}_{00}(r_{12})(r_{12}) \tag{11}$$

The central term and the zero range pseudo-potential one, which represents the effects of single nucleon knock-on exchange, are given, respectively, as

$$\begin{aligned} v_{00}(r_{12}) &= \left[ 7999 \frac{\exp(-4r_{12})}{4r_{12}} - 2134 \frac{\exp(-2.5r_{12})}{2.5r_{12}} \right] \quad MeV, \\ \hat{j}_{00}(r_{12}) &\approx -276 \left[ 1 - 0.005 \left( \frac{E}{A_e} \right) \right] \quad MeV \, fm^3. \end{aligned} \tag{12}$$

The energy of the emitted light particle is corrected for recoil, $E = A_d Q/(A_e + A_d)$. Many studies had been used the M3Y-Reid interaction to investigate the α-decay process [6,12]. For a spherical-deformed interacting pair, the double-folding potential is calculated numerically using the multipole expansion method [31,32]. The matter (charge) density distributions of the deformed daughters are considered in the two-parameter Fermi [33] form with half-density radius, $R(\theta) = R_0 \left( 1 + \beta_2 Y_{20}(\theta) + \beta_3 Y_{30}(\theta) + \beta_4 Y_{40}(\theta) + \beta_6 Y_{60}(\theta) \right)$. Here, $\theta$ is the angle made by the direction of the position vector $\vec{r}$ with respect to the symmetry axis of the deformed daughter nucleus. $\beta_{i=2,3,4,6}$ are the quadrupole, octupole, hexadecapole, and hexacontatetrapole deformation parameters, respectively. For spherical daughter nuclei, this radius will be independent of orientation. The half-density radius of the density distribution is used as $R_0 = 1.28 A^{1/3} + 0.8 A^{-1/3} - 0.76$ [34] where $A$ is the mass number of the mentioned nucleus. The diffuseness parameter is taken as $a=0.54$. The density distribution of the spherical α-particle is used in its standard Gaussian form [35].

### III. Results and discussion

We consider α decays from 496 heavy and superheavy α-emitters. They divided into four groups of 166 even (Z)-even (N), 117 odd-even, 141 even-odd, and 72 odd-odd emitters. The half-life times of the studied decays are taken according to the most recent data in the NUBASE2012 evaluation of nuclear properties [2]. We also considered the recent information about the ground-state spin and parity [2] of the involved nuclei, their atomic masses, and the corresponding values of the released energy (Q-values), according to the Atomic Mass Evaluation AME2012 [36]. We have excluded the decays of the unmeasured half-life times and intensities, but have estimated values from trends in neighbor nuclei. We excluded also the decays of Q-values estimated from systematic trends with large uncertainty, $\Delta Q > 0.070$ MeV [36], and the unfavored decays with no information on the ground-state spin-parity of any of the involved nuclei. The chosen α emitters have atomic and mass numbers ranges of Z = 52– 118 and A = 105– 294, respectively. Our calculations are performed in the framework of the DDCM. The different deformation components [37] of daughter nuclei, up to the hexacontatetrapole deformation parameter, are taken into account. In our calculations, we adopt the angle-averaging of the individual assault frequency ($\nu(\theta)$) and penetration probability ($P_{WKB}(\theta)$) over the different emission angles, because we are interested in the detailed behavior of these two quantities with the mass number of the α emitters. The averaging of the whole decay width ($\Gamma(\theta)=\hbar \nu(\theta)P(\theta)$) is more suitable for the studies concerning the behavior of the half-life time or the preformation probability. For the α decay of $^{252}$Cf, as an example, minor differences are obtained in the values of the average decay width and the half-life time calculated by the two averaging methods. The α-decay from ground-state parent to ground-state daughter of the same spin-parity is a favored transition. The angular momentum carried by the emitted α-particle in this case is zero. For odd-A and odd-odd nuclei, the angular momentum ($\ell$) transferred by the α-particle cannot be clearly determined for the cases where the daughter and parent have different ground state spin-parity. In such cases, we adopt the smallest one ($\ell_{min}$) among all allowed $\ell$ values. This is a reasonable procedure because in most cases the α-particle prefers to carry the minimum allowed angular momentum. A measure of the agreement between the calculated half-life times and the experimental values for studied nuclei is the standard deviation defined as,

$$\sigma = \sqrt{\sum_{i=1}^{N}[log_{10}(T_i^{cal}/T_i^{exp})]^2/(N-1)}.$$

Figure (1) shows the variation of the decimal logarithm of the assault frequency ($log_{10}\nu$), for the preformed α cluster inside all studied parent nuclei, with the parent mass number ($A_P$). In this figure, a gentle linear behavior of $log_{10}\nu$ as a function of $A_P$ is obtained. This behavior can be well described by the linear expression,

$$log_{10}\nu(A) = -0.00035 A_P + 21.53036$$

Because of this gentle variation, one would use $log_{10}\nu$ as an average value overall the studied cases. The averaging process yields, $log_{10}\nu = 21.45777$. When the daughter nuclei are assumed to be spherical in the calculations, $log_{10}\nu$ is slightly changed. This indicates that the decimal logarithm of the assault frequency weakly depends on the deformation of the daughter nuclei.

Another comparison between the phenomenological calculations of the α-decay universal curve and our realistic calculations is shown in Fig. 2. This figure presents the ratio of the decimal logarithm of the WKB penetration probability, Eq. (6), based on the M3Y-Reid NN interaction, to the same quantity calculated by Eq. (2) for the considered decays in the present work. The calculated ratio is presented versus the mass number of the parent nuclei. The results show that this ratio for the decays involving even-even nuclei is ranged between 0.983 and 1.027, with an average value of 1.004. The calculated ratio for the group of odd-even (even-odd) radioactive nuclei is ranged between 0.979 and 1.091 (0.980 and 1.089), with an average value of 1.016 (1.014). For the decay processes involving odd-odd nuclei, the ratio between the two mentioned quantities is ranged between 0.991 and 1.086, with an average value of 1.019. As can be seen, the minimum (maximum) ratio of the WKB penetration probability to the approximated one is obtained for the decays involving even-even (odd-odd) nuclei.

Now, we plot in Figs. 3 and 4 the suggested universal curves for α decay, Eq.(4). They are based on the WKB penetration probability that calculated in terms of the M3Y-Reid NN interaction. In these figures, the decimal logarithm of the experimental half-lives, $log_{10} T_{1/2}^{Exp}(s)$, are plotted versus the two quantities,

$$-log_{10} P_{WKB} + C = -log_{10} P_{WKB} - 21.61694 \qquad (13)$$

and

$$-log_{10} P_{WKB} + C_A = -log_{10} P_{WKB} - 21.68953 + 0.00035 A_P, \qquad (14)$$

respectively. Here, the parameters $C$ and $C_A$ are given in terms of the two values of the assault frequency obtained from Fig. 1 as $C = log_{10}(\ln 2) - log_{10} \nu = -21.61694$ and $C_A = log_{10}(\ln 2) - log_{10} \nu (A) = -21.68953 + 0.00035 A_P$, respectively.

The results are classified in the different panels of Figs. 3 and 4 according to the mentioned four groups of the decays involving even-even, odd-even, even-odd, and odd-odd nuclei. It is clearly seen that the results are well described by straight lines on the chosen double logarithmic scale of Figs. 3 and 4. We list in Table I the parameterization of the suggested universal curves based on our calculations, in terms of constant average assault frequency. According to Eq.(4), the equations of the obtained straight lines give an individual average value for the preformation probability inside each presented group of nuclei. Upon using the constant assault frequency, the fitting straight lines presented in Fig. 3 yield preformation probability values of 0.0647, 0.0325, 0.0254, and 0.0145 for forming the α cluster inside even-even, odd-even, even-odd and odd-odd radioactive nuclei, respectively. When using an $A_p$-dependent assault frequency, Fig. 4, we obtain average values of 0.0647, 0.0325, 0.0255, and 0.0144, for the α preformation probability inside the same mentioned groups, respectively. However, the obtained average values of the preformation probability from the two representations in Figs. 3 and 4 are almost the same. Such differences in the preformation probability values for the different studied groups are due to many influences. One of these influences is due to the confirmed pairing effect [21]. Another effect is the difference between the ground state spin-parity of parent and formed daughter nuclei, in the unfavored decays of the nuclei involving odd number(s) of nucleons. Actually, our results for the average values of the preformation probability confirm the results obtained previously in Ref. [21], where the smallest (largest) preformation probability was assigned for the odd-odd (even-even) α emitters. Also, our results confirm that the preformation probability inside the even-odd α emitters, which have unpaired neutron, is less than it in the neighboring odd-even emitters of the same shell and subshell closures but have unpaired proton instead [21]. The difference between the extracted average values of the preformation probability in these two cases is a little bit larger in our results. The reason is mainly due to the increase in the number of considered unfavored decays with non-zero transferred angular momentum for the group of even-odd α emitters (76 decays) than it for the group of odd-even emitters (59 decays). The effect of pairing and that of non-zero transferred angular momentum appear in the single universal curve calculations of Poenaru *et al*. [18,24,25] as a fitting hindrance factor $h_U$. However, to get an effective preformation probability including both the effect of unpaired nucleons and that of the spin-parity difference in the unfavored decays, we may add the value of the hindrance factor ($h_U$) [25] to that of $-log_{10} S$ in Eq. (4), $log_{10} \acute{S} = log_{10} S - h_U$. In this way, the values of $log_{10} S$ and $h_U$ given in Ref. [25] yield an effective preformation probability with values of 0.0150, 0.0067, 0.0048, and 0.0018 for even-even, odd-even, even-odd and odd-odd nuclei, respectively. The differences between these values confirm again our results and the obtained results in Ref. [21], regarding the pairing effect.

In Fig. 3, the standard deviation of the half-life times obtained according to the suggested universal curves, from the experimental data, has a minimum (maximum) value of σ = 0.318 (0.830) for the even-even (odd-odd) group. The standard deviation values for the odd-even and even-odd groupsare 0.711 and 0.700, respectively.

Upon considering the $A_p$-dependent assault frequency instead of the constant one, Fig. 4, a slight improvement in the standard deviation is obtained. In this case, the standard deviations for the even-even, odd-even, even-odd, and odd-odd groups become 0.311, 0.706, 0.694, and 0.824, respectively. For these mentioned groups, and considering the same recent experimental half-lives [2] and Q-values [36] used here, the parameterization of the universal curve given in Ref. [25] (with a hindrance factor $h_U$) yields standard deviation values of σ = 0.389, 0.764, 0.813 and 0.914, respectively. Actually, the mentioned parameterization of Ref. [25] is given upon fitting to the experimental half-lives and Q-values used in it. According to the data used therein [25], standard deviation values of 0.354, 0.565, 0.640 and 0.826 are obtained by its authors for the studied groups of even-even, odd-even, even-odd, and odd-odd nuclei.

Finally, to obtain a single universal curve for the α decay half-lives of radioactive nuclei, we plot in Fig. 5 the decimal logarithm of the experimental half-lives versus the sum of $-log_{10} P_{WKB}$ and the parameter $C(S, \nu)$, where

$$C(S,\nu) = log_{10}(\ln 2) - log_{10} \nu - log_{10} S = -21.61694 - log_{10} S.$$

This parameter, $C(S,\nu)$, includes the contributions of both the preformation probability and the assault frequency, as a constant average value, Table I. It includes also the hindrance factor $h_U$ that usually used in the universal curve calculations to consider the effect of unpaired nucleons. The values of $C(S,\nu)$ for the classified four groups are given in Table I. According to this parameterization, a standard deviation of σ = 0.625 between the calculated half-lives based on the suggested single universal curve and the experimental ones is obtained for all considered nuclei. However, upon calculating the WKB penetration probability, the suggested universal curve would help in estimating the unknown half-lives of α decay mode for the radioactive heavy nuclei as well as the synthesized super-heavy elements.

## IV. **SUMMARY AND CONCLUSION**

We investigated the validity of using semi-microscopic calculations of the α-particle penetration probability and assault frequency, based on the WKB approximation and realistic NN interaction, to obtain universal curves for the half-lives of α-radioactive nuclei. The nuclear and Coulomb parts of the interaction potential between the α and daughter clusters are computed using the semi-microscopic double-folding model, based on the realistic M3Y-Reid NN interaction for its real nuclear part. The deformations of daughter nuclei are taken into account. We carried out the calculations considering four groups of 166 even (Z)-even (N), 117 odd-even, 141 even-odd, and 72 odd-odd heavy parent nuclei. We found that the decimal logarithm of the obtained assault frequency is well described as a function of the parent mass number by a linear behavior with a gentle slope. This gentle variation allowed us either to use the $A_P$-dependence of $log_{10}\nu$ or to be approximated by an average constant value. We next plotted the decimal logarithm of the experimental half-lives versus the sum of $-log_{10} P_{WKB}$ plus the parameter $C = -21.61694$, and plus $C_A = -21.68953 + 0.00035 A_P$ instead, individually. C and $C_A$ are obtained based on the deduced average value of $log_{10}\nu$ and based on its $A_P$-dependence, respectively. Accordingly, we obtained straight lines that can be used as universal α-decay curves for the different groups of even-even, odd-even, even-odd, and odd-odd α emitters. Upon using average assault frequency, the presented universal curves in Fig. 3 yielded preformation probability values of 0.0647, 0.0325, 0.0254, and 0.0145 for forming the α cluster inside the mentioned groups of nuclei, respectively. The differences in the deduced values of preformation probability inside the classified groups refer to both the pairing effect and the difference in the ground-state spins and/or parities of parents and formed daughter nuclei, in the unfavored decays odd-A and odd-odd nuclei. Considering the $A_p$-dependent assault frequency instead of the constant one yields almost the same values of the preformation probability and improves slightly the standard deviations of the given universal curves. A single universal curve for the α decay is finally obtained by plotting the decimal logarithm of the experimental half-lives versus the sum of $-log_{10}P_{WKB}$ plus the parameter $C(S,\nu) = -21.61694 - log_{10}S$. The number of parameters is then reduced where both the preformation probability and the average assault frequency, in addition to the hindrance pairing factor, are all included in this parameter $C(S,\nu)$. The values of $C(S,\nu)$ are given for the classified four groups. However, in addition to using recent reported experimental half-life times and Q-values data in the fitting procedure, the present parameterization of the α single universal curve has two advantages. First, it is based on semi-realistic calculations. Second, it has less number of parameters.

---------------------------------------------------

[1] E-mails: wseif@sci.cu.edu.eg and wseif@yahoo.com .

**Figures and Tables Captions:**

**Fig. 1**: The decimal logarithm of the assault frequency for the 496 studied decays versus the mass number of parent nuclei.

**Fig. 2**: The ratio between the decimal logarithm of the WKB penetrability calculated by Eq. (6), in terms of the M3Y-Reid NN interaction, and that calculated by the approximate expression of Eq. (2) for the 496 studied decays as a function of the mass number of parent nuclei.

**Fig. 3**: Universal curves for α decay based on the WKB penetration probability in terms of the M3Y-Reid NN interaction. The decimal logarithm of the experimental half-life times are plotted versus the sum of $-log_{10} P_{WKB}$ plus the parameter $C = log_{10}(\ln 2) - log_{10} \nu = -21.61694$, which is given in terms or assault frequency deduced from Fig. 1 as an average constant value. The four panels show the results for the different groups of even-even, odd-even, even-odd, and odd-odd α emitters.

**Fig. 4**: Same as Fig. 3 but the parameter $C_A = log_{10}(\ln 2) - log_{10} \nu(A) = -21.68953 + 0.00035 A_P$, is given in terms of the assault frequency deduced from Fig. 1 as a function of $A_P$.

**Fig. 5**: Single universal curve for α decay based on the WKB penetration probability, in terms of the M3Y-Reid NN interaction. The parameter $C(S,\nu) = log_{10}(\ln 2) - log_{10}\nu - log_{10} S = -21.61695 - log_{10} S$, which added to the negative of the decimal logarithm of the WKB penetration probability on the x-axis is given in terms of the assault frequency deduced from Fig. 1, as an average constant value, and the values of $S$ as obtained from Fig. 3. The values $C(S,\nu)$ are given in Table I.

**Table I**: The parameterization of the suggested universal curves of α decay as extracted from the realistic calculations presented in Figs. 1 and 3, considering the assumption of constant assault frequency. Also presented are the values of the parameter $C(S,\nu) = log_{10}(\ln 2) - log_{10}\nu - log_{10} S$ for the suggested single universal curve shown in Fig. 5. The standard deviations of the calculated half-life times based on the suggested universal curves relative to the experimental ones are shown.

**Table I**:

| Quantity | Group | Value |
|---|---|---|
| Assault frequency ($\nu$) | All | $log_{10}\nu = 21.45777$ |
| Preformation probability (S) | even-even | 0.0647 |
| | odd-even | 0.0325 |
| | even-odd | 0.0254 |
| | odd-odd | 0.0145 |
| Standard deviation (σ) | even-even | 0.318 |
| | odd-even | 0.711 |
| | even-odd | 0.700 |
| | odd-odd | 0.830 |
| $C(S,\nu)$ | even-even | -20.428 |
| | odd-even | -20.129 |
| | even-odd | -20.022 |
| | odd-odd | -19.778 |
| Overall standard deviation | All | 0.625 |

**Fig. 1**

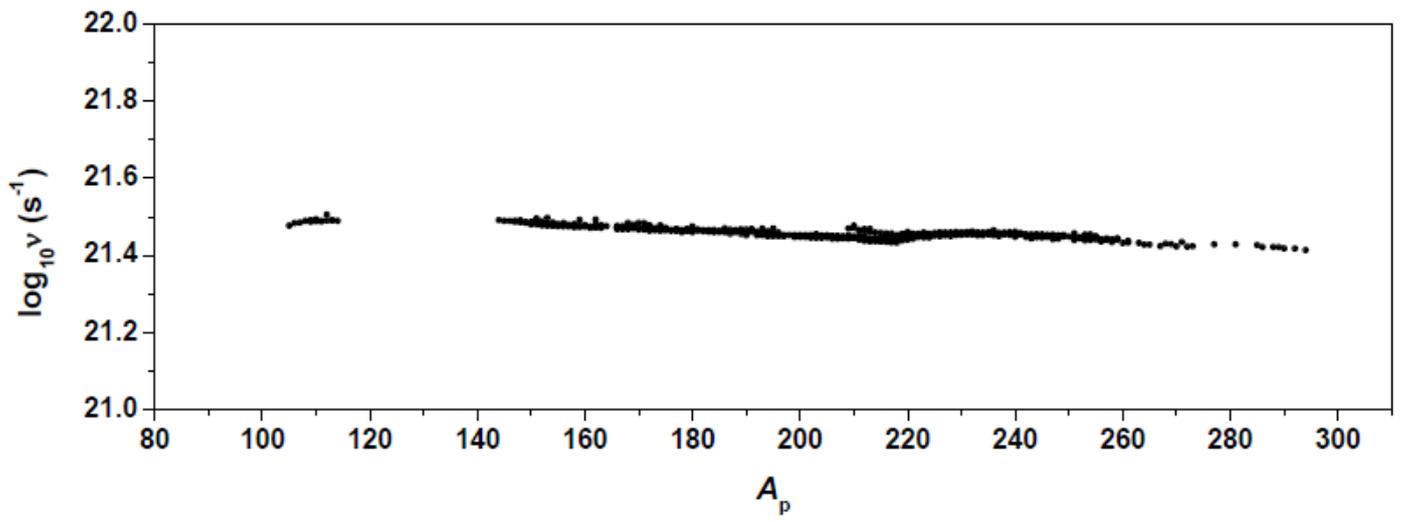

**Fig. 2**

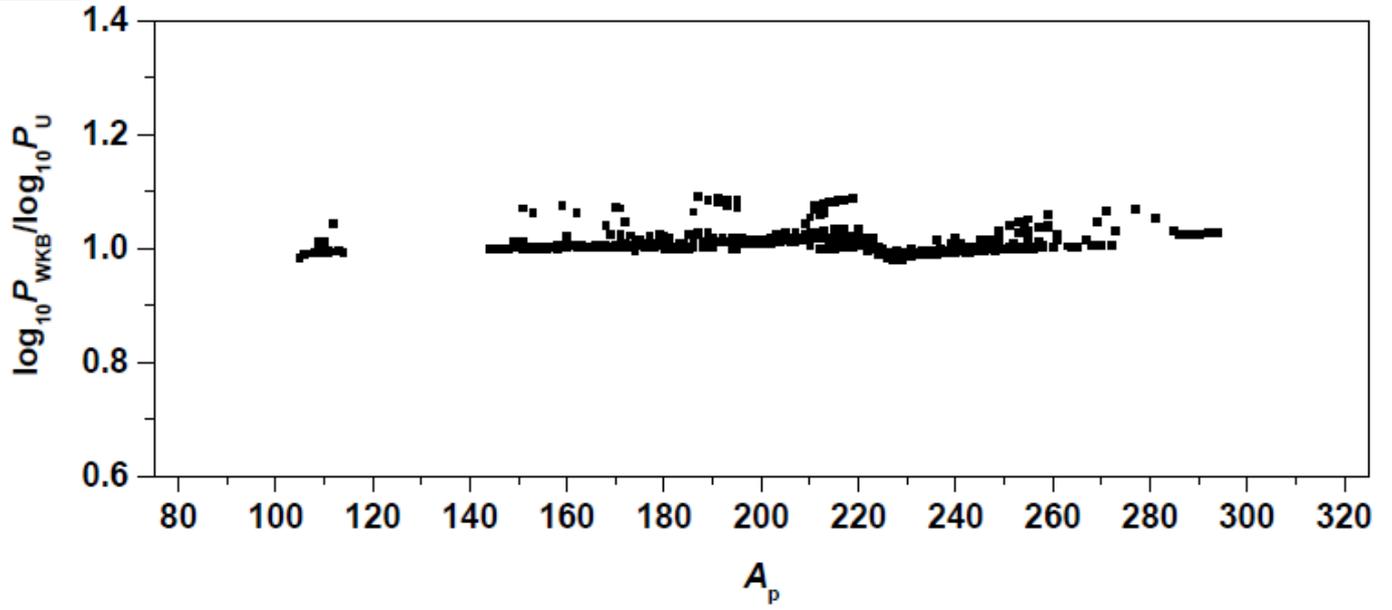

**Fig. 3**

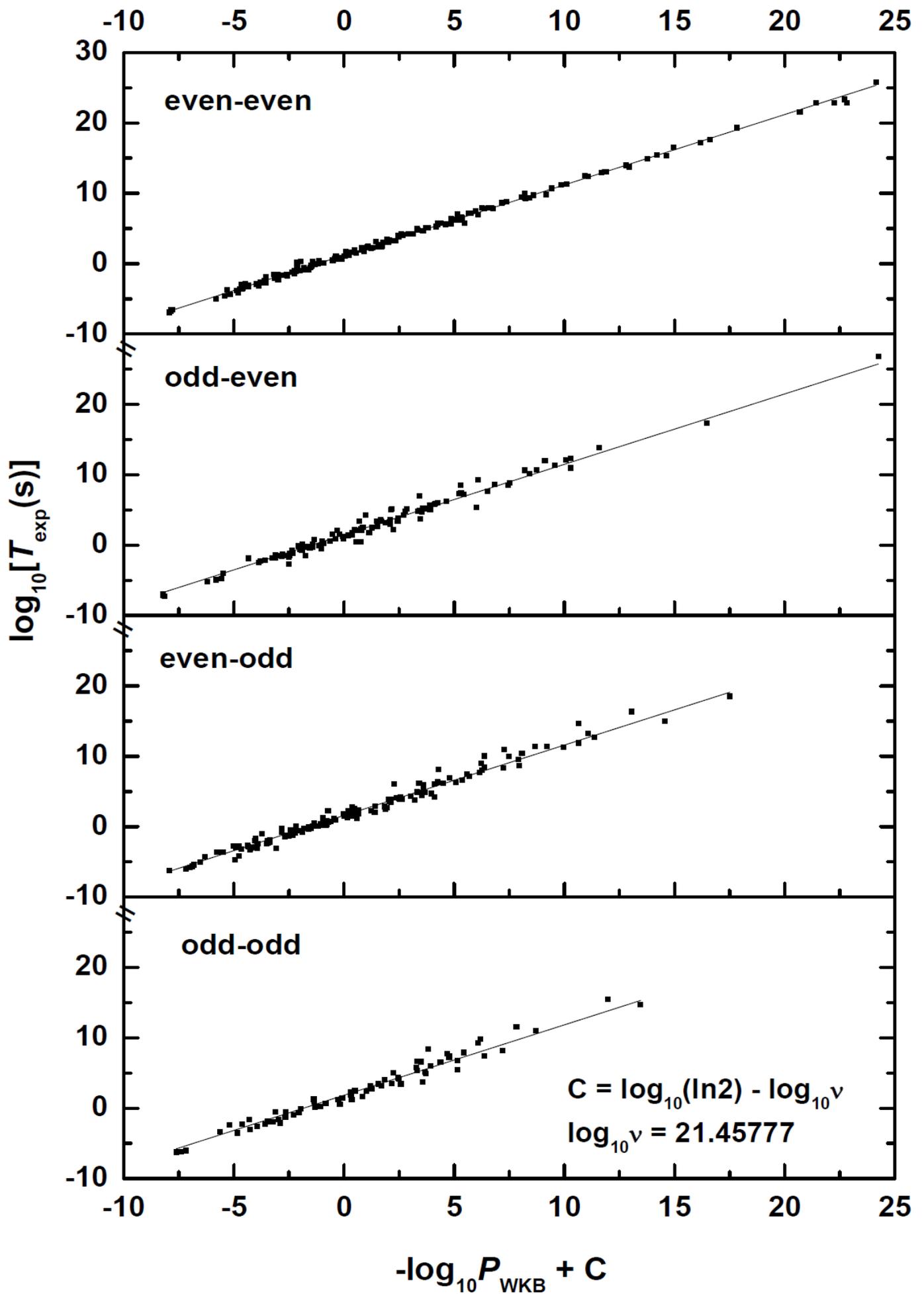

**Fig. 4**

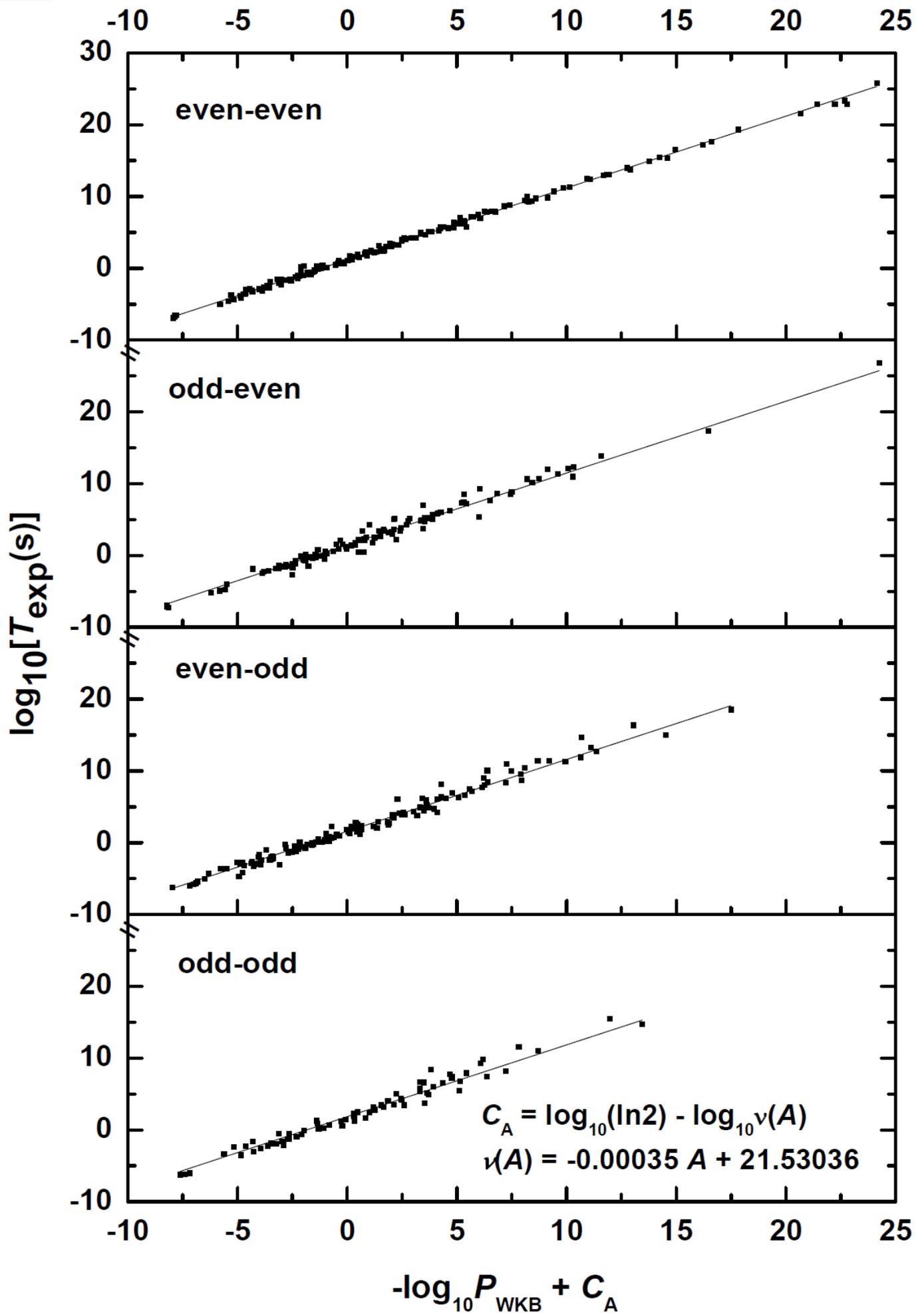

**Fig. 5**

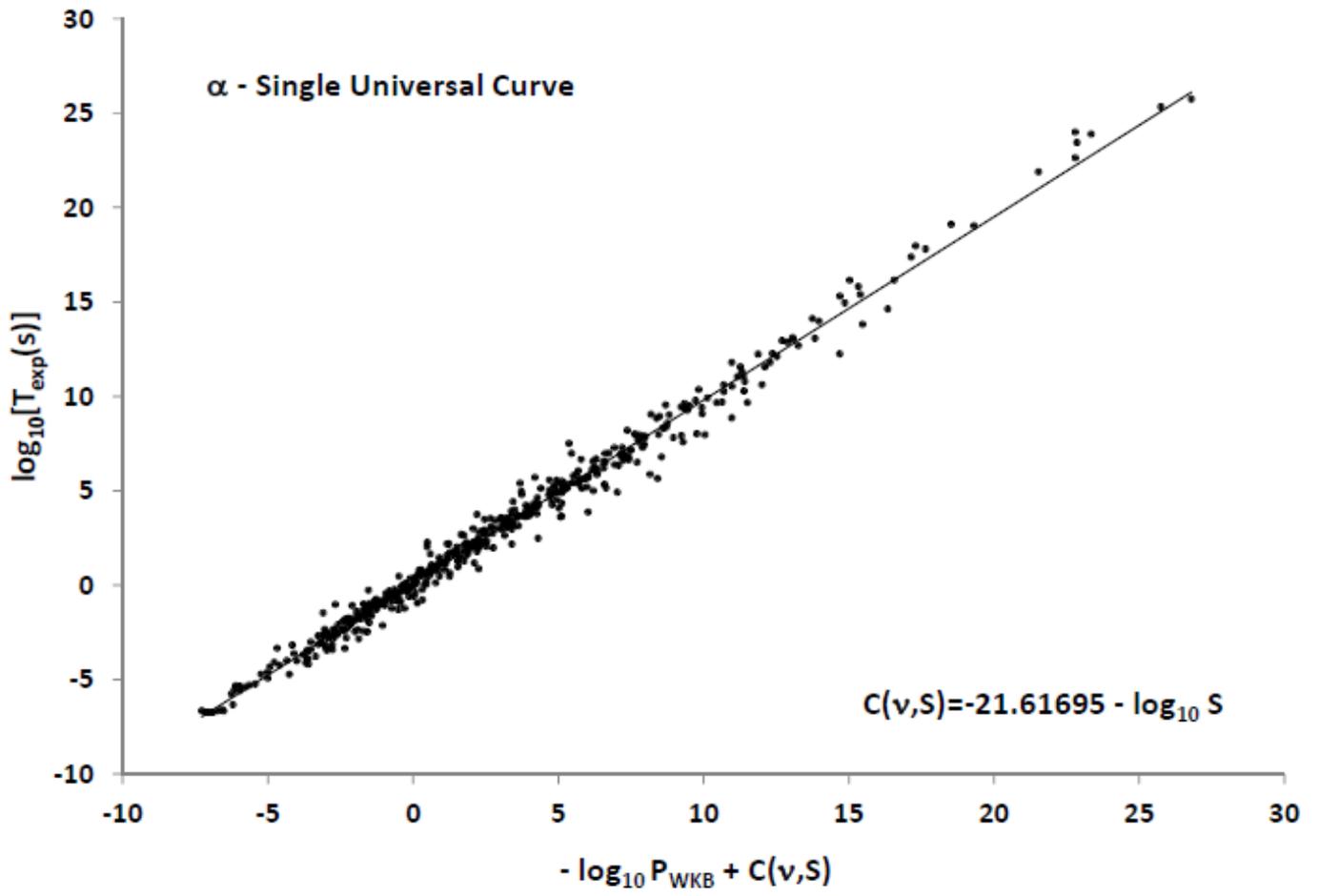